\documentclass[aps,prb,reprint,amsmath,amssymb,amsfonts,twocolumn,superscriptaddress,footinbib]{revtex4-2}

\usepackage[utf8]{inputenc}
\usepackage[T1]{fontenc} 

\usepackage{hyperref}
\usepackage{graphicx}
\usepackage{textcomp}
\usepackage{xcolor}
\usepackage{color}
\usepackage{bm}
\usepackage{physics}
\usepackage{xspace}
\usepackage{multirow}
\usepackage{changes}
\usepackage{dsfont}

\newcommand{\Tc}{\ensuremath{T_{\mathrm{c}}}\xspace}
\newcommand{\e}{\mathrm{e}}
\newcommand{\pr}{^{\prime}}

\newcommand{\cno}[1]{c^{ }_{#1}}
\newcommand{\cdag}[1]{c^{\dagger}_{#1}}

\newcommand{\NaCoOHO}{Na$_x$CoO$_2\cdot y$H$_2$O\xspace}

\begin{document}

\preprint{}

\title{Efficient fluctuation exchange approach to low-temperature spin fluctuations and superconductivity: from the Hubbard model to \NaCoOHO}

\author{Niklas Witt}
	\email{niwitt@uni-bremen.de}
	\affiliation{Institut f{\"u}r Theoretische Physik, Universit{\"a}t Bremen, Otto-Hahn-Allee 1, 28359 Bremen, Germany
	}
	\affiliation{Bremen Center for Computational Materials Science, Universit{\"a}t Bremen, Am Fallturm 1a, 28359 Bremen, Germany
	}

\author{Erik G. C. P. van Loon}
	\affiliation{Institut f{\"u}r Theoretische Physik, Universit{\"a}t Bremen, Otto-Hahn-Allee 1, 28359 Bremen, Germany
	}
	\affiliation{Bremen Center for Computational Materials Science, Universit{\"a}t Bremen, Am Fallturm 1a, 28359 Bremen, Germany
	}

\author{Takuya Nomoto}
	\affiliation{
	Department of Applied Physics, The University of Tokyo,7-3-1 Hongo, Bunkyo-ku, Tokyo 113-8656, Japan
	}

\author{Ryotaro Arita}
	\affiliation{
	Department of Applied Physics, The University of Tokyo,7-3-1 Hongo, Bunkyo-ku, Tokyo 113-8656, Japan
	}
	\affiliation{ 
	RIKEN Center for Emergent Matter Science, 2-1 Hirosawa, Wako, Saitama 351-0198, Japan 
	}

\author{Tim O. Wehling}
	\affiliation{Institut f{\"u}r Theoretische Physik, Universit{\"a}t Bremen, Otto-Hahn-Allee 1, 28359 Bremen, Germany
	}
	\affiliation{Bremen Center for Computational Materials Science, Universit{\"a}t Bremen, Am Fallturm 1a, 28359 Bremen, Germany
	}

\date{\today}

\begin{abstract}
Superconductivity arises mostly at energy and temperature scales that are much smaller than the typical bare electronic energies. Since the computational effort of diagrammatic many-body techniques increases with the number of required Matsubara frequencies and thus with the inverse temperature, phase transitions that occur at low temperatures are typically hard to address numerically. In this work, we implement a fluctuation exchange (FLEX) approach to spin fluctuations and superconductivity using the "intermediate representation basis" (IR) [Shinaoka et al., PRB 96, 2017] for Matsubara Green functions. This FLEX+IR approach is numerically very efficient and enables us to reach temperatures on the order of $10^{-4}$ in units of the electronic band width in multi-orbital systems. After benchmarking the method in the doped repulsive Hubbard model on the square lattice, we study the possibility of spin-fluctuation-mediated superconductivity in the hydrated sodium cobalt material \NaCoOHO reaching the scale of the experimental transition temperature \(\Tc=4.5\)~K and below.
\end{abstract}


\maketitle

\section{Introduction} 
The physics of unconventional superconductivity has been a longstanding problem in condensed matter physics. Over the course of decades many different systems have been discovered, like heavy fermion compounds \cite{Steglich1979,White2015}, cuprates \cite{Bednorz1986,Keimer2015}, iron-based superconductors \cite{Kamihara2006,Stewart2011}, twisted two-dimensional (2D) materials \cite{Cao2018,Balents2020}, and infinite-layer nickelate \cite{Li2019}.

Finding a microscopical description for these materials is a difficult task since correlations as well as complexity need to be accounted for appropriately. The inherent complexity of real materials arises from the interplay of many internal degrees of freedom and typically covering multiple energy scales. For instance, screening of the Coulomb interaction often involves electronic bands reaching up to 100~eV in energy. On the other side, superconductivity emerges when thermal energies are on a scale of 10~meV for \(\Tc\) cuprate systems down to a few 10~{\textmu}eV in several heavy fermion systems. Hence, four or even more orders of magnitude of electronic energies are typically involved in the electronic structure of superconducting materials. For the theoretical modeling this has practical consequences. Distinct energy scales require large but accurate frequency grid sampling and processing. This frequently limits the phase space that can be studied by diagrammatic many-body methods.

One particular material example for this complex interplay of different degrees of freedom and energy scales is given by the water intercalated sodium cobalt oxide, \NaCoOHO, which features superconductivity with transition temperatures reaching \(\Tc=4.5\)~K \cite{Takada2003}. This material consists of layered cobalt oxide planes being separated by sodium ions and water molecules. The Co atoms are arranged on a triangular lattice and hole doped, rendering it a possible realization of a resonating-valence-bond state, related to high-temperature superconductivity \cite{Anderson1973,Anderson1987}. However, until now neither an experimental nor a theoretical consensus has been reached on the origin of the superconducting pairing.

Theoretically proposed pairing types include a spin triplet \(p\)\,- or \(f\)-wave driven by ferromagnetic fluctuations \cite{Kuroki2004,Mochizuki2005,Mazin2005,Ogata2007}, a spin singlet extended \(s\)-wave \cite{Mochizuki2006,Kuroki2006}, a chiral \(d\)+\(id\)\,-\,wave \cite{Baskaran2003,Kiesel2013}, an odd frequency gap \cite{Johannes2004,Mazin2005}, or conventional phonon-assisted \(s\)-wave pairing \cite{Yada2006,Yada2008}. For each of them, experimental results can be found that support or deny their realization \cite{Sakurai2015}, making the analysis quite delicate. This controversy about the pairing type originates from several problems. They include a general instability of the \NaCoOHO compound due to water evaporation with an accompanying large dependence on sample conditions \cite{Sakurai2015}. On the theoretical side, a multi-orbital model is necessary to accurately describe the electronic structure \cite{Yanase2005}, which makes computational studies very challenging. It might be one of the reasons why no microscopic studies of the superconducting instability have been reported on the temperature scale of \(\Tc \sim 5\,\)K.

In this work, we implement a fluctuation exchange (FLEX) approach \cite{Bickers1989a,Bickers1989b,Arita2000,Takimoto2004,Mochizuki2005,Yamazaki2020,Sakakibara2020,Bjoernson2021} using the intermediate representation (IR) basis \cite{Shinaoka2017,Chikano2018,Chikano2019,Otsuki2020} and study the possibility of spin-fluctuation-mediated superconductivity in \NaCoOHO. The IR basis provides a compact representation of imaginary-time quantities that additionally enables the usage of sparsely sampled data grids \cite{Li2020}. As a result, the numerical cost of calculations can be considerably reduced permitting, e.g., new ab initio approaches \cite{Wang2020}. Here, we use this combined FLEX+IR approach to perform calculations at very low temperatures. We study the magnetic properties of \NaCoOHO and investigate the possibility of  triplet superconductivity occurring on the scale of the experimental \(\Tc\). 

The remainder of this work is structured as follows: In Sec. \ref{sec:Methods} we will briefly review the FLEX approximation and explain the application of the IR basis. To illustrate the accuracy and efficiency of our approach, we first show benchmark studies on the single-orbital Hubbard model in Sec. \ref{sec:Hubbard_benchmark}. Subsequently, we use our method to research the possibility of spin-fluctuation-driven superconductivity in the \NaCoOHO system at very low temperatures. For this, we study the Fermi surface, filling and interaction dependence of the spin susceptibility and superconducting instability in Sec. \ref{sec:NaCoO2}.

\section{Methods}\label{sec:Methods}

\subsection{Fluctuation exchange approximation}
The FLEX approximation introduced by Bickers et al. \cite{Bickers1989a,Bickers1989b} is a perturbative diagrammatic approach that treats spin and charge fluctuations self-consistently. It can be derived from a Luttinger-Ward functional \cite{Luttinger1960} containing an infinite series of closed bubble and ladder diagrams. As such it is a conserving approximation \cite{Baym1961,Baym1962}. Due to its perturbative nature, FLEX cannot sufficiently capture strong coupling physics but it performs well in the weak-coupling regime. It is suitable for studying systems with strong spin fluctuations in Fermi liquids and near quantum critical points.

In this paper we employ the multi-orbital extension of FLEX \cite{Takimoto2004,Kubo2007} for which we consider the (antisymmetrized) local interaction Hamiltonian
\begin{align}
	H_{\mathrm{int}} = \frac{1}{4}\sum_i \sum_{\xi_1\xi_2\xi_3\xi_4} \Gamma^{0}_{\xi_1\xi_4,\xi_3\xi_2}\cdag{i\xi_1}\cdag{i\xi_2}\cno{i\xi_3}\cno{i\xi_4}
\end{align}
where the operators \(\cdag{i\xi}\) (\(\cno{i\xi}\)) create (destroy) an electron at site \(i\) in a state \(\xi=(l,\sigma)\), which is a combined orbital and spin index. The bare vertex \(\Gamma^0\) is expressed as
\begin{align}
	\begin{split}
		\Gamma^{0}_{\xi_1\xi_4,\xi_3\xi_2} = &-\frac{1}{2} U^{\mathrm{S}}_{l_1l_4,l_3l_2} \bm{\sigma}_{\sigma_1\sigma_4}\cdot\bm{\sigma}_{\sigma_2\sigma_3}\\
		&+ \frac{1}{2} U^{\mathrm{C}}_{l_1l_4,l_3l_2} \delta_{\sigma_1\sigma_4}\delta_{\sigma_2\sigma_3}
	\end{split}
\end{align}
with the interaction matrices
\begin{align*}
	U^{\mathrm{S}}_{ij,kl} = \left\{
	\begin{aligned}
		&U\\ &U\pr\\ &J \\ &J\pr
	\end{aligned}\right.\;,\quad
	U^{\mathrm{C}}_{ij,kl} = \left\{
	\begin{aligned}
		&U &\mathrm{if}\, i = j = k = l\\
		&-U\pr + 2J &\mathrm{if}\, i = k \neq l = j\\
		&2U\pr -J &\mathrm{if}\, i = j \neq l = k\\
		&J\pr &\mathrm{if}\, i = l \neq k = j
	\end{aligned}\right.
\end{align*}
where \(U\) and \(U\pr\) are the local intra- and inter-orbital interactions, \(J\) is the inter-orbital exchange interaction or Hund's coupling, and \(J\pr\) is the pair-hopping between two orbitals. Due to symmetry, they are related by \(U = U\pr + J + J\pr\) and \(J=J\pr\).

In FLEX, the self-energy can be calculated from
\begin{align}
	\Sigma_{lm}(k) = \frac{T}{N}\sum_{q}\sum_{l\pr,m\pr} V_{ll\pr,mm\pr}(q)G_{l\pr m\pr}(k-q)
	\label{eq:FLEX_multi_orb_Sigma_formula}
\end{align}
where \(k = (i\omega_n,\bm{k})\) and \(q=(i\nu_m,\bm{q})\) denote crystal momentum and Matsubara frequencies \(\omega_n=(2n+1)\pi T\) (\(\nu_m = 2m\pi T\)) for fermions (bosons), \(T\) is the temperature, and \(N\) is the number of sites. The interaction consists of contributions from the spin and charge channel as
\begin{align}
	\begin{split}
		V(q) &= \frac{3}{2}U^{\mathrm{S}}\qty[\chi^{\mathrm{S}}(q)- \frac{1}{2}\chi^0(q)]U^{\mathrm{S}} + \frac{3}{2}U^{\mathrm{S}}\\
		&+ \frac{1}{2}U^{\mathrm{C}}\qty[\chi^{\mathrm{C}}(q)- \frac{1}{2}\chi^0(q)]U^{\mathrm{C}}  - \frac{1}{2}U^{\mathrm{C}}\;.
	\end{split}
\label{eq:FLEX_multi_orb_normal_V_formula}
\end{align}
The charge and spin susceptibility entering Eq. (\ref{eq:FLEX_multi_orb_normal_V_formula}) are defined by
\begin{align}
	\chi^{\mathrm{C}}(q) = \frac{\chi^0(q)}{\mathds{1} + \chi^0(q)U^{\mathrm{C}}}\quad,\quad
	\chi^{\mathrm{S}}{}(q) = \frac{ \chi^0(q)}{\mathds{1} - \chi^0(q)U^{\mathrm{S}}}\quad.
	\label{eq:Charge_spin_susceptibility}
\end{align}
with the unity operator \(\mathds{1}\) and the irreducible susceptibility
\begin{align}
	\chi^0_{ll\pr,mm\pr}(q) = - \frac{T}{N}\sum_{k} G_{lm}(k+q)G_{m\pr l\pr}(k)\,.
	\label{eq:Irreducible_susceptibility}
\end{align}
We use Eqs. (\ref{eq:FLEX_multi_orb_Sigma_formula})-(\ref{eq:Irreducible_susceptibility}) to self-consistently solve the Dyson equation
\begin{align}
	G(k)^{-1} = G^0(k)^{-1} - \Sigma(k)
\end{align}
with the bare Green function given by
\begin{align}
	G^0(i\omega_n,\bm{k}) = \frac{\mathds{1}}{i\omega_n\mathds{1} -(H^0(\bm{k}) -\mu\mathds{1})}\,. 
	\label{eq:bare_gf}
\end{align}
$H^0(\bm{k})$ is the non-interacting Hamiltonian and $\mu$ denotes the chemical potential which needs to be adjusted in every iteration to keep the electron density $n$ fixed. The fractions in Eqs. (\ref{eq:Charge_spin_susceptibility}) and (\ref{eq:bare_gf}) are to be understood as inversions.
	
In the presence of strong magnetic fluctuations, it is possible to study the superconducting phase transition within FLEX. For this purpose, we consider the linearized Eliashberg theory with the gap equation reading
\begin{align}
	\lambda \Delta^{\eta}_{lm}(k) = \frac{T}{N}\sum_{q}\sum_{l\pr,m\pr} V^{\eta}_{ll\pr,m\pr m}(q)F^{\eta}_{l\pr m\pr}(k-q)\;.
	\label{eq:FLEX_multi_orb_Delta_formula}
\end{align}
It is diagonal in the spin singlet- and triplet-pairing channel (\(\eta\)\,=\,s,\,t) with the anomalous Green function \mbox{\(F^{\eta}(k) = -G(k)\Delta^{\eta}(k)G^{\mathrm{T}}(-k)\)} and respective interactions
\begin{align}
	V^{\mathrm{s}}(q) &= \frac{3}{2}U^{\mathrm{S}}\chi^{\mathrm{S}}(q)U^{\mathrm{S}} - \frac{1}{2}U^{\mathrm{C}}\chi^{\mathrm{C}}(q)U^{\mathrm{C}}  + \frac{1}{4}\qty(3U^{\mathrm{S}} + U^{\mathrm{C}})\,,\notag\\
	V^{\mathrm{t}}(q) &= -\frac{1}{2}U^{\mathrm{S}}\chi^{\mathrm{S}}(q)U^{\mathrm{S}} - \frac{1}{2}U^{\mathrm{C}}\chi^{\mathrm{C}}(q)U^{\mathrm{C}}  - \frac{1}{4}\qty(U^{\mathrm{S}} - U^{\mathrm{C}})\,.
	\label{eq:FLEX_multi_orb_triplet_interaction}
\end{align}
\(\lambda\) and \(\Delta\) in the linearized gap equation (\ref{eq:FLEX_multi_orb_Delta_formula}) represent an eigenvalue and eigenvector of the Bethe-Salpeter equation, respectively \cite{Bickers1989a,Bulut1992}. We solve this eigenvalue problem by using the power iteration method. The superconducting transition temperature is found if the eigenvalue \(\lambda\) reaches unity.

\subsection{Intermediate representation basis}
The basic objects of diagrammatic many-body methods like FLEX are Green functions and derived quantities which are computed numerically on finite imaginary-time and Matsubara frequency grids. Using conventional uniform grids to represent Green functions, calculations require grid sizes that increase linearly with inverse temperature upon cooling the system. In practice, this prohibits calculations at low temperatures as the required amount of data becomes too large to be stored or processed. One of several approaches \cite{Pao1994,Takeuchi2018,Schrodi2019} to tackle these problems is to use a compact representation of Green functions as given by (orthogonal) continuous basis functions, like Legendre polynomials \cite{Boehnke2011,Dong2020}, Chebyshev polynomials \cite{Gull2018} or numerical basis functions \cite{Kaltak2020}. 

The IR basis \cite{Shinaoka2017,Chikano2018,Chikano2019,Otsuki2020} is such an orthogonal numerical basis in which Green functions can be efficiently and compactly represented. The basis functions are defined by the singular value expansion of the kernel that connects Green function and spectral function \footnote{The additional factor \(\omega\) in the bosonic case prevents divergence for \(\omega\to0\). The spectral function needs to be redefined as well, see Ref. [\onlinecite{Shinaoka2017}].}: 
\begin{align}
	K^{\alpha}(\tau,\omega) = \omega^{\delta_{\alpha,\mathrm{B}}}\frac{\e^{-\omega\tau}}{1\pm\e^{-\beta\tau}} = \sum_{l=0}^{\infty}S^{\alpha}_l \;\! U_l^{\alpha}(\tau)V_l^{\alpha}(\omega)
	\label{eq:IR_basis_definition}
\end{align}
Here, \(\{U^{\alpha}_l(\tau)\},\,\{V^{\alpha}_l(\omega)\}\) denote the IR basis functions and \(S_l^{\alpha}\) are the exponentially decaying singular values. The expansion is uniquely defined by fermionic or bosonic statistics \(\alpha\in\qty{\mathrm{F},\,\mathrm{B}}\) and a dimensionless parameter \mbox{\(\Lambda = \beta\omega_{\mathrm{max}}\)}, where \(\beta\) is the inverse temperature and \(\omega_{\mathrm{max}}\) a cutoff frequency which captures the spectral width of the system.

\begin{figure}[t]
	\includegraphics[width=0.48\textwidth]{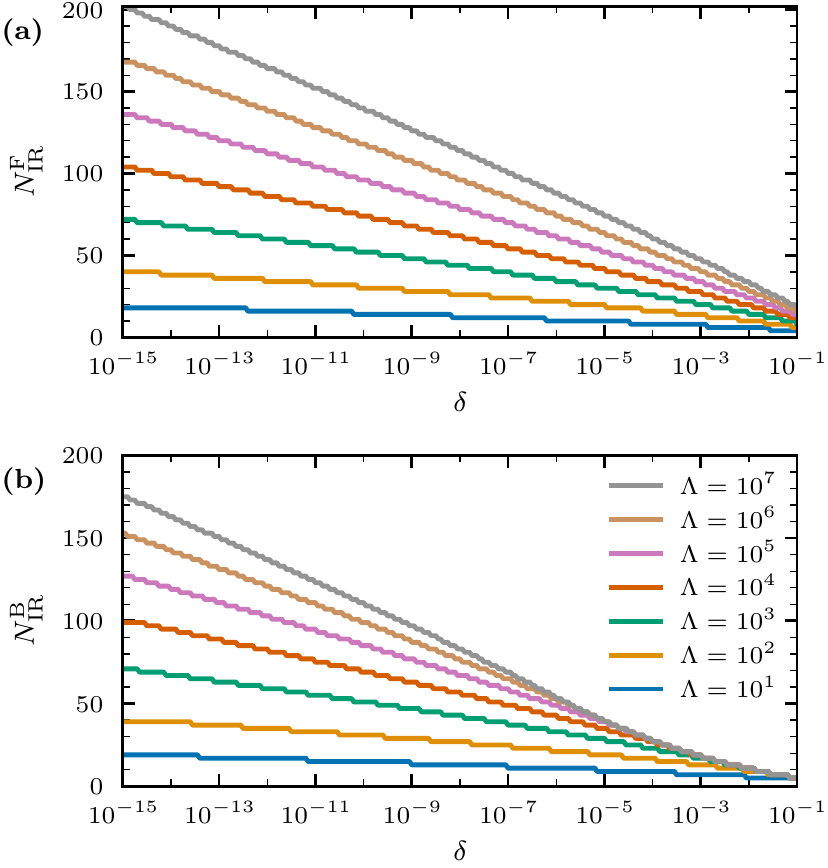}
	\caption{Number of IR basis functions \(N_{\mathrm{IR}}^{\alpha}\) needed to sufficiently expand (a) fermionic or (b) bosonic Green functions within an error bound \(\delta\). The imaginary-time and Matsubara frequency grid sizes can be chosen equally large.}
	\label{fig:IR_basis_size_accuracy}
\end{figure}

The representation within the IR basis provides a controlled way to store Green functions. This means that the truncation error \(\delta\) of the expansion
\begin{align}
	G^{\alpha}(x) = \sum_{l=0}^{l_{\mathrm{max}}}G^{\alpha}_l U^{\alpha}_l(x)\qquad [x=\tau,\,i\omega_n]
\end{align}
with \(N_{\mathrm{IR}}^{\alpha}=l_{\mathrm{max}}+1\) basis functions is controllable. It is determined from the singular values by \(\delta \leq S_{l_{\mathrm{max}}}/S_0\). Due to their exponential decay, only a small number \(N_{\mathrm{IR}}^{\alpha}\) is necessary to compactly represent Green function data with high accuracy as is shown in Fig. \ref{fig:IR_basis_size_accuracy}. Compared to other basis sets like Legendre or Chebyshev polynomials, the IR basis holds a superior compactness, especially at low temperatures.

Another advantage of the IR basis is the possibility to generate sparse imaginary-time and frequency grids with a size equal to the number of basis functions \cite{Li2020}. This scheme ideally requires an even (odd) number for fermionic (bosonic) quantities because of which the lines in Fig. \ref{fig:IR_basis_size_accuracy} are step functions. The sparse grids offer the benefit of decreased data storage while performing intermediate steps of solving diagrammatic calculations efficiently, like computing Fourier transformation by simple matrix multiplications \footnote{For a technical comparison of the IR basis based sparse sampling approach and other methods see Table 1 of Ref. [\onlinecite{Dong2020}].}.
 
In practical calculations, a desired accuracy \(\delta\) is chosen and \(\Lambda\) is set such that \(\Lambda\geq\beta\omega_{\mathrm{max}}\) holds for a fixed \(\omega_{\mathrm{max}}\). Then, the \(\{U^{\alpha}_l\}\) functions are precomputed on imaginary-time and Matsubara frequency grids. The evaluation of Eq. (\ref{eq:IR_basis_definition}) is numerically expensive. However, the open-source \texttt{irbasis} software package \cite{Chikano2019} provides numerical basis functions as solutions to Eq. (\ref{eq:IR_basis_definition}) which can be quickly accessed and implemented easily. Throughout this paper we employed \(\Lambda=10^4\) and \mbox{\(\delta=10^{-8}\)} which corresponds to small basis and grid sizes of \(N^{\mathrm{F}}_{\mathrm{IR}}=62\) and \(N^{\mathrm{B}}_{\mathrm{IR}}=57\).

\section{Benchmark: Single-orbital square lattice Hubbard model}\label{sec:Hubbard_benchmark} 
The Hubbard model is a fundamental model used to study correlated electron physics, particularly the interplay of magnetism and unconventional superconductivity. Despite its simplicity it captures many essential physics important to interacting quantum systems. Thus, a multitude of many-body approaches has been developed to simulate the properties of the Hubbard model \cite{LeBlanc2015,Schaefer2020}. Therefore, it constitutes an excellent system to benchmark our FLEX+IR approach to former FLEX and further studies of magnetism and superconductivity. 

In this regard, we consider the repulsive single-orbital Hubbard model on a square lattice, which also serves as a relevant study case for cuprates \cite{Keimer2015}. Taking into account nearest and next-nearest-neighbor hoppings \(t\) and \(t\pr\), the single-particle dispersion is given by
\begin{align}
\varepsilon_{\bm{k}} = 2t\qty[\cos(k_x)+\cos(k_y)]+4t\pr\cos(k_x)\cos(k_y)\;.
\end{align}
In the following \(t\) is the unit of energy. We set the local interaction to an intermediate value of \(U/t=4\). For an assessment of the performance of the FLEX+IR method introduced, here we first compare to an earlier FLEX work of one of the current authors \cite{Arita2000}. To this end, we adapted the \(N=64^2\) lattice sites in our calculations and replaced the uniform 2048 Matsubara frequency grid of Ref. [\onlinecite{Arita2000}] with the IR basis sampling.

The Hubbard model contains different magnetic fluctuations whose relative strength can be controlled by the Fermi surface shape, i.e., by changing \(t\pr\) and the electron filling \(n\). To contemplate different physical situations, we inspect the possibility of both dominant antiferromagnetism (AF) and ferromagnetism (F) by using the parameters \(t\pr/t=0,\,n=0.85\) and \(t\pr/t=0.5,\,n=0.3\), respectively.

\begin{figure}[t]
	\includegraphics[width=0.48\textwidth]{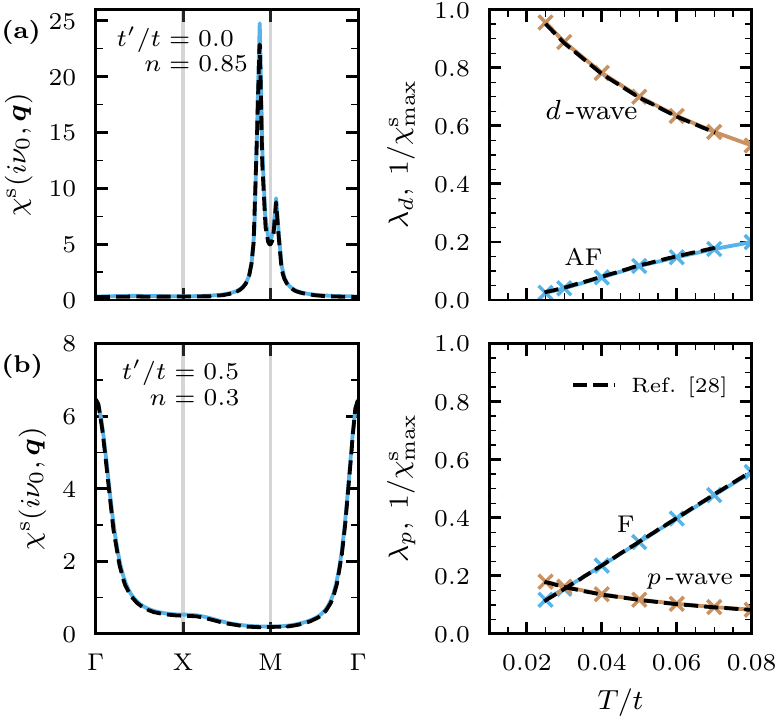}
	\caption{Comparison of static spin susceptibility (left column) at \(T/t=0.03\) and eigenvalue of the Eliashberg equation as well as inverse magnetic susceptibility at the leading instability (right column) as calculated with our FLEX+IR implementation with results from Ref. [\onlinecite{Arita2000}] (dashed lines). The rows show two different situations with dominant (a) antiferromagnetism (AF: \(t\pr/t=0\),\,\(n=0.85\)) and (b) ferromagnetism (F: \(t\pr/t=0.5\),\,\(n=0.3\)) for \(U/t=4\).}
	\label{fig:Hubbard_Literature_comparison}
\end{figure}

First we examined the spin susceptibility \(\chi^{\mathrm{s}}\). The results for the static spin susceptibility \(\chi^{\mathrm{s}}(i\nu_0=0,\bm{q})\) are shown along high-symmetry paths in the Brillouin zone in the left column of Fig. \ref{fig:Hubbard_Literature_comparison}. For a direct comparison, we also included the results of Ref. [\onlinecite{Arita2000}]. Clearly, the agreement between both data sets is excellent. The dominant structures and magnitude of the incommensurate antiferromagnetic and the weaker ferromagnetic fluctuations are reproduced exactly. 

The presence of strong magnetic fluctuations can drive unconventional superconductivity. To study its appearance, we calculated the superconducting eigenvalue \(\lambda\). In the case of dominant AF fluctuations, we consider a singlet-pairing gap with \(d_{x^2-y^2}\equiv d\)-symmetry while we choose the degenerate triplet \(p\)\,-wave state for dominant F fluctuations. Their respective eigenvalues are shown in the right column of Fig. \ref{fig:Hubbard_Literature_comparison} together with the inverse of \(\chi^{\mathrm{s}}(0,\bm{Q})\) at the wave vector \(\bm{Q}\) of the leading instability, which signifies magnetic ordering. As can be seen, the AF fluctuations are strong enough to enable \(d\)-wave superconductivity with a \(\Tc\approx0.02t\), whereas the \(p\)\,-wave solution is not realized. This is mainly due to stronger self-energy renormalization for \(t\pr>0\) and a smaller prefactor in the triplet-pairing potential \(V^{\mathrm{t}}(q)\) in Eq.  (\ref{eq:FLEX_multi_orb_triplet_interaction}). Once again, we included data from Ref. [\onlinecite{Arita2000}], which agree very well. This demonstrates that by employing the IR basis we can reduce the necessary frequency points by a factor of \(\sim33\) while achieving the same results under persistent accuracy.

In a second step, we use our FLEX+IR approach to study the superconducting and magnetic phase diagram of the square lattice Hubbard model with \(t\pr/t=0\). An additional comparison to numerical methods beyond FLEX will be made.

\begin{figure}[t]
	\includegraphics[width=0.48\textwidth]{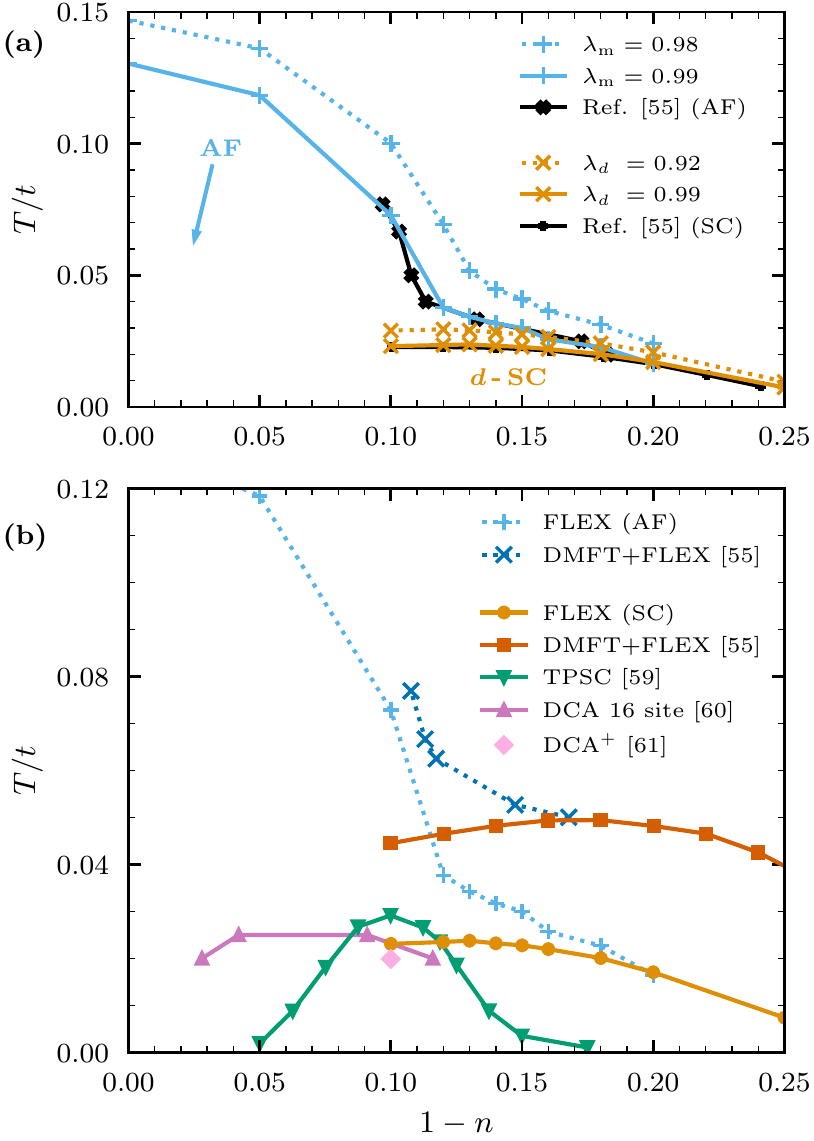}
	\caption{Phase diagram of the Hubbard model with \(t\pr/t=0\) and \(U/t=4\). (a) Comparison of different magnetic (AF) and superconducting (SC) eigenvalues calculated by the FLEX+IR approach with results from Ref. [\onlinecite{Kitatani2015}]. (b) Comparison of calculated phase boundaries from FLEX+IR to a variety of methods including DMFT+FLEX \cite{Kitatani2015}, TPSC \cite{Kyung2003}, DCA on a 16 site cluster \cite{Gull2013}, and DCA\(^+\) \cite{Staar2014}.}
	\label{fig:Hubbard_model_phase_diagram}
\end{figure}

In order to map out the phase diagram, we performed calculations for different fillings and temperatures. Regions of strong magnetic and superconducting fluctuations can be identified by analyzing and extrapolating the corresponding magnetic (\(\lambda_{\mathrm{m}} = U\chi^{0}_{\mathrm{max}}\)) and superconducting (\(\lambda_d\)) eigenvalues. In Fig. \ref{fig:Hubbard_model_phase_diagram}(a) we show the \(n\)\,-\,\(T\) diagram for two extracted values of \(\lambda_{\mathrm{m}}\) and \(\lambda_d\) to indicate the evolution of the phase boundaries for \(\lambda\to1\). Additionally, we included independent FLEX results by Kitatani et al. \cite{Kitatani2015} (\(\lambda_{\mathrm{m},d}=0.99\)) to verify our accuracy. This latter comparison yields an excellent agreement.

The results show that \(\Tc\) grows monotonically with the electron filling with some flattening of the curve around a hole doping of 0.15 as has been reported previously \cite{Yanase2003}. We cannot, however, make a statement about the underdoped region near half-filling due to strong AF fluctuations preventing the FLEX cycle from converging. This can be seen from \(\lambda_{\mathrm{m}}\to1\), which masks the superconducting domain below 0.1 hole doping. This issue is inherently a part of the theory due to the diverging denominator of the spin susceptibility. Here, the strong fluctuations result from better nesting conditions on the Fermi surface with less doping which becomes even more profound for larger \(U\).

At this point, we should comment on the designation of phase boundaries at finite temperatures, since the Mermin-Wagner theorem \cite{Mermin1966,Hohenberg1967} actually prohibits the formation of (perfect) long-range ordered phases associated with spontaneous breaking of continuous symmetries at finite temperatures in two dimensions. The results shown here are best understood in the context of quasi-two-dimensional systems: It has been shown that purely two-dimensional systems show very similar results to quasi-two-dimensional systems with a weak but finite three-dimensional character as long as the out-of-plane coherence length is large \cite{Arita2000}.

In Fig. \ref{fig:Hubbard_model_phase_diagram}(b) we compare our phase diagram obtained from FLEX for \(\lambda_{\mathrm{m},d}=0.99\) with phase diagrams reported in the literature which have been calculated using DMFT+FLEX (Dynamical Mean-Field) \cite{Kitatani2015}, Two-Particle Self-Consistency (TPSC) \cite{Kyung2003}, Diagrammatic Cluster Approximation (DCA) on a 16 site cluster \cite{Gull2013}, and DCA\(^+\) \cite{Staar2014}. On a qualitative level all approaches under consideration yield maximally achievable superconducting critical temperatures on the same order of magnitude. Also the shape of the phase boundary of the AF region agrees between FLEX and FLEX+DMFT.

On a close, more quantitative level, however, there are profound differences between the phase diagrams revealed by the different methods: The most prevalent difference between all methods lies in the structures of the superconducting dome in the phase diagram. The filling dependence of this dome shape varies significantly. Due to the reasons of the previous discussion, FLEX does not establish this dome structure. It can be retrieved by incorporating strong correlation effects as contained in DMFT, DCA,  and also in TPSC. The level at which correlations are incorporated, however, strongly influences the exact doping dependence.

\begin{figure*}[t]
	\includegraphics[width=0.95\textwidth]{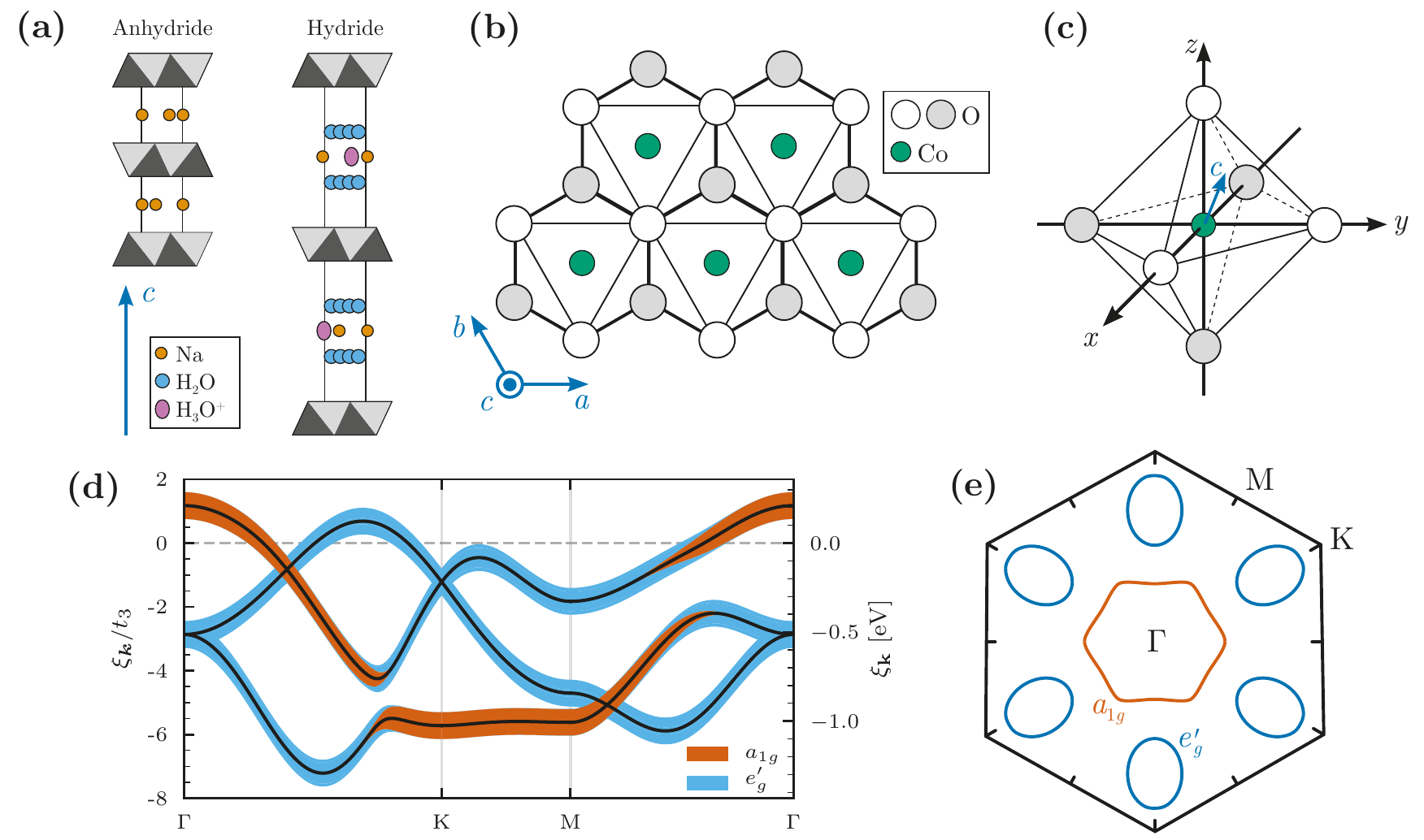}
	\caption{Crystal and electronic structure of the \NaCoOHO compound. (a) Vertical layered structure of CoO\(_2\) planes (light and dark gray) with intercalated Na\(^{+}\), H\(_2\)O, and H\(_3\)O\(^+\). (b) Top view on CoO\(_2\) planes showing triangular sublattice of Co ions with surrounding O ions. (c) CoO\(_6\) octahedron which is trigonally deformed by the layered structure. (d) Electronic band structure with orbital character projections indicated by surrounding color patches. Model details are given in Appendix \ref{sec:app_NaxCoO_TB}. Energies are measured with respect to the chemical potential \(\xi_{\bm{k}}=\varepsilon_{\bm{k}}-\mu\). (e) Fermi surface corresponding to the band structure of panel (d).}
	\label{fig:NaCoO_system}
\end{figure*}

\section{Sodium cobalt oxide}\label{sec:NaCoO2}
The pairing type of superconductivity and its interplay with magnetism in \NaCoOHO is a very controversial issue as we have elucidated in the Introduction. In the following, we apply the FLEX+IR approach to study this problem.

\subsection{Crystal and electronic structure}\label{sec:NaCoO2_structure}
\NaCoOHO is commonly synthesized by soft-chemical methods from the parent compound Na$_{0.7}$CoO$_2$. The latter is a layered material consisting of cobalt oxide planes which are separated by sodium ions, c.f. Fig. \ref{fig:NaCoO_system}(a).  The CoO\(_2\) planes are composed of edge-shared CoO\(_6\) octahedras that place the Co ions on a perfect triangular lattice as depicted in Figs. \ref{fig:NaCoO_system}(b) and \ref{fig:NaCoO_system}(c). During hydration, water molecules and hydronium ions are intercalated between the CoO\(_2\) planes. As a consequence, the separation between the CoO\(_2\) planes in the \(c\)-direction increases while the CoO\(_6\) octahedra contract in that direction. The material becomes thus more anisotropic, i.e., H\(_2\)O intercalation enhances two-dimensionality in the CoO\(_2\) planes. 

The Co atoms have partially filled \(t_{2g}\) bands which are electron doped by the Na ions. In the simplest approximation, their filling is \(n=5+x\) where \(x\) is the Na content. Upon Na doping, a rich phase diagram \cite{Foo2004,Sakurai2015} with weak correlations for low dopings (\(x\sim0.3\)) and strong correlations for high dopings (\(x\sim0.7\)) emerges. In this phase characterization, the superconducting region is placed around \(x\approx0.3\). However, this classification had been made without consideration of possible additional doping from the H\(_3\)O\(^+\) ions because their presence was only discovered at a later time \cite{Sakurai2015}. Due to this, the filling of the \(t_{2g}\)-bands might be larger in the superconducting phase, locating it in the strongly correlated region \cite{Ogata2007,Wilhelm2015}.

To model the electronic structure, we use a three-band tight-binding model for the \(t_{2g}\) bands as formulated by Mochizuki et al. \cite{Mochizuki2005} which describes the low-energy characteristics of LDA band structure calculations \cite{Singh2000} quite well. This model includes a crystal field term accounting for the trigonal deformation of the CoO\(_6\) octahedra because of the plane height reduction. It leads to a splitting of the \(t_{2g}\) orbitals into a higher \(a_{1g}\) and lower twofold \(e_g\pr\) levels. The exact details on this model are presented in Appendix \ref{sec:app_NaxCoO_TB}. The corresponding band structure is shown orbitally resolved for a Co valence of \(s=3.645\) or respective electron filling of \(n=5.355\) in Fig. \ref{fig:NaCoO_system}(d). Panel (e) contains the associated Fermi surface. It consists of one large \(a_{1g}\) hole pocket around the Brillouin zone center and six elliptically shaped \(e_{g}\pr\) hole pockets near the K points. The latter play an important role in creating strong ferromagnetic fluctuations since they have a large density of states and offer good nesting conditions for \(\bm{Q}\approx(0,0)\) \cite{Yanase2005,Mochizuki2005,Johannes2004}.

There has been much discussion on the actual existence of the \(e_g\pr\) pockets on the Fermi surface in the literature. It stems from the fact that ARPES measurements \cite{Hasan2004,Yang2004,Yang2005,Shimojima2006,Geck2007} locate them below the Fermi level. However, these results might be due to surface effects \cite{Pillay2008} since PES \cite{Kubota2004} and Shubnikov-de Haas measurements \cite{Oeschler2008} seem to support their existence. Theoretical studies showed that the \(e_g\pr\) pockets are suppressed in charge self-consistent LDA+DMFT calculations \cite{Boehnke2014}, while LDA+DFMT performed with a realistic Hund's coupling \(J\) can stabilize them \cite{Ishida2005}. The problem of locating the \(e_g\pr\) pockets in energy is very delicate since variations in the crystal field splitting (layer height), electron filling, and band width renormalization influence the Fermiology of \NaCoOHO. In this work, we study the interplay of spin fluctuations and superconductivity for different models of the Fermi surface. We start with the type of Fermi surface considered also in Ref. [\onlinecite{Mochizuki2005}] and vary the Fermi surface shape and topology afterward.

\begin{figure}[b]
	\includegraphics[width=0.48\textwidth]{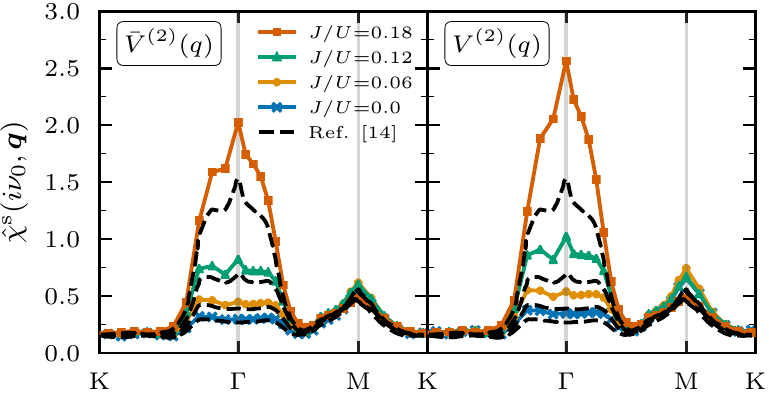}
	\caption{Comparison of largest eigenvalue of the static spin susceptibility to results from Ref. [\onlinecite{Mochizuki2005}] at \(T/t_3=0.02\) using a \(32\times32\) \(\bm{k}\)-mesh. The second-order correction used in the calculations is different between both panels (see text and Eqs. (\ref{eq:Second_order_V_wrong}) and (\ref{eq:Second_order_V_correct})).}
	\label{fig:NaCoO_literature_spin_susceptibility}
\end{figure}

\begin{figure*}[t]
	\centering
	\includegraphics[width=0.85\textwidth]{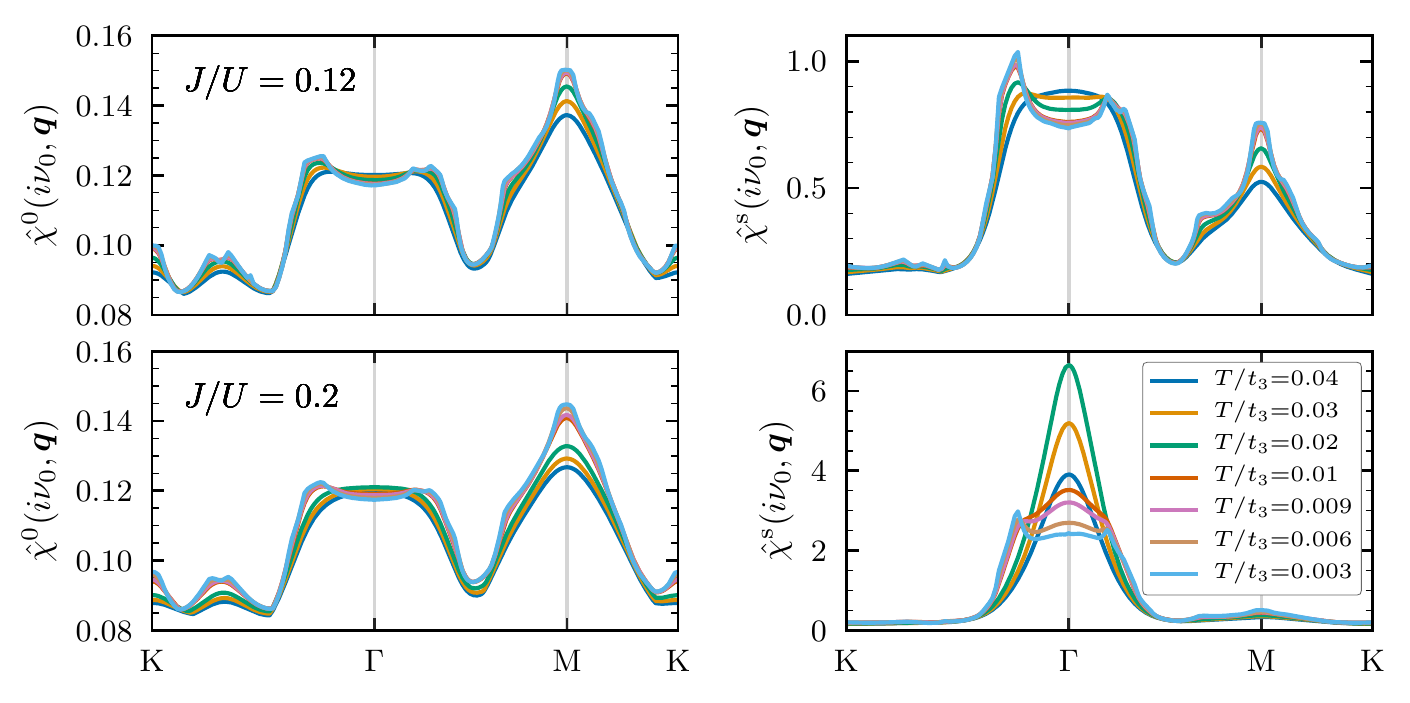}
	\caption{Evolution of the largest eigenvalues of static (\(i\nu_0=0\))  irreducible susceptibility \(\hat{\chi}^0\) and spin susceptibility \(\hat{\chi}^{\mathrm{s}}\) for two exchange interactions \(J/U\).}
	\label{fig:NaCoO_irr_spin_susceptibility_over_T}
\end{figure*}

To this end, we first reproduce the results given in Ref. [\onlinecite{Mochizuki2005}] within FLEX+IR and then extend the calculations to lower temperatures. We adapted the interaction strength of \(U=6\) in units of the hopping \(t_3\) (\(U \sim 1.1\)~eV, see appendix \ref{sec:app_NaxCoO_TB}) and we vary the Hund's coupling \(J\) as a ratio of \(U\). For the initial comparison, we use a \(\bm{k}\)-mesh of \(32\times32\) as in Ref. [\onlinecite{Mochizuki2005}], but the low-temperature calculations demand a denser grid sampling for which we found \(N_{\bm{k}}=210^2\) lattice sites to be converged.

\subsection{Spin Susceptibilities}
To check the accuracy of our implementation, we calculated the static spin susceptibility and compare our results to Ref. [\onlinecite{Mochizuki2005}], where calculations were carried out for a temperature of \(T/t_3=0.02\) and different \(J/U\) values. It should be noted that a different second-order correction \(\bar{V}^{(2)}(q)\) to the FLEX interaction has been employed in Ref. [\onlinecite{Mochizuki2005}] which is given by
\begin{align}
	\bar{V}^{(2)}(q) &= - \frac{1}{4}(U^{\mathrm{S}}+U^{\mathrm{C}})\chi^0(q)(U^{\mathrm{S}}+U^{\mathrm{C}})\,.
	\label{eq:Second_order_V_wrong}
\end{align}
Comparing it to the second-order contribution from Refs. [\onlinecite{Yada2005,Kubo2007,Sknepnek2009,Yanagi2010,Horvath2011}] as implemented in our code
\begin{align}
	V^{(2)}(q) &= - \frac{3}{4}U^{\mathrm{S}}\chi^0(q)U^{\mathrm{S}}-\frac{1}{4}U^{\mathrm{C}}\chi^0(q)U^{\mathrm{C}}
	\label{eq:Second_order_V_correct}
\end{align}
it becomes evident that \(\bar{V}^{(2)}(q)\) incorrectly includes mixing between spin and charge channel contributions. In Fig. \ref{fig:NaCoO_literature_spin_susceptibility} we show the largest eigenvalue of the static spin susceptibility \(\hat{\chi}^{\mathrm{s}}\)
for both interactions together with data by Mochizuki et al. from Ref. [\onlinecite{Mochizuki2005}]. It can be seen that the results are very well reproduced if \(\bar{V}^{(2)}(q)\) is implemented (left panel). Comparing it to the implementation of \(V^{(2)}\) (right panel) shows that the incorrect mixing of fluctuation channels leads to a reduction of fluctuation strength.

Generally, the system contains F as well as AF fluctuations. By increasing \(J\), ferromagnetism is strongly enhanced while the AF fluctuations are slightly decreased. The latter are generated by scattering on the \(a_{1g}\) surface as well as between different \(e_g\pr\) pockets, whereas the F fluctuations emerge mostly from intra-pocket scattering in the \(e_g\pr\) sheets. The charge fluctuations are negligibly small and not shown here.

The previously discussed results were at a relatively high temperature of \(T/t_3=0.02\), which corresponds to \(\sim\)50~K. In order to properly understand the superconducting transition, a lower temperature range on the order of the experimental critical temperatures needs to be investigated. In Fig. \ref{fig:NaCoO_irr_spin_susceptibility_over_T} we show the temperature evolution of the largest eigenvalues of static irreducible susceptibility \(\hat{\chi}^0\) and spin susceptibility \(\hat{\chi}^{\mathrm{s}}\) for two exchange interaction ratios \(J/U\). \(\hat{\chi}^0\) does not show a strong dependence on \(T\). The peak at the M point becomes slightly enhanced while the structure around the \(\Gamma\) point changes a bit.

Contrary to this, \(\hat{\chi}^{\mathrm{s}}\) shows a strong \(T\) dependence. By cooling the system, the ferromagnetic fluctuation strength exhibits a non-monotonous behavior with a strong enhancement of the peak at \(\bm{Q}=(0,0)\) for \mbox{\(T/t_3\approx0.02\)}. This non-monotonous evolution traces back to an almost divergent \(\hat{\chi}^{\mathrm{s}}\) stemming from the denominator in Eq. (\ref{eq:Charge_spin_susceptibility}) approaching zero. In other words, the simulated case is close of a ferromagnetic instability. Indeed, we could not converge calculations for larger Hund's couplings \(J/U\geq0.22\) since \(J\) favors the formation of ferromagnetic order. For \(J/U=0.2\) we find that the maximum in \(\hat{\chi}^{\mathrm{s}}\) jumps at some intermediate temperature \(T\sim 0.01\) from having an absolute maximum at \(\bm{Q}=(0,0)\) to an absolute maximum at finite \(\bm{q}\)-vectors. Hence, some long-wavelength spin waves are the favored type of fluctuation in this regime. The \(\bm{q}\)-vectors associated with these spin waves match well to the minor and major axes of the \(e_g\pr\) pockets.  Since the Fermi surface becomes less thermally smeared out, the scattering between opposite edges is favored.

\subsection{Triplet superconductivity possible?}
\begin{figure}[t]
	\centering
	\includegraphics[width=0.48\textwidth]{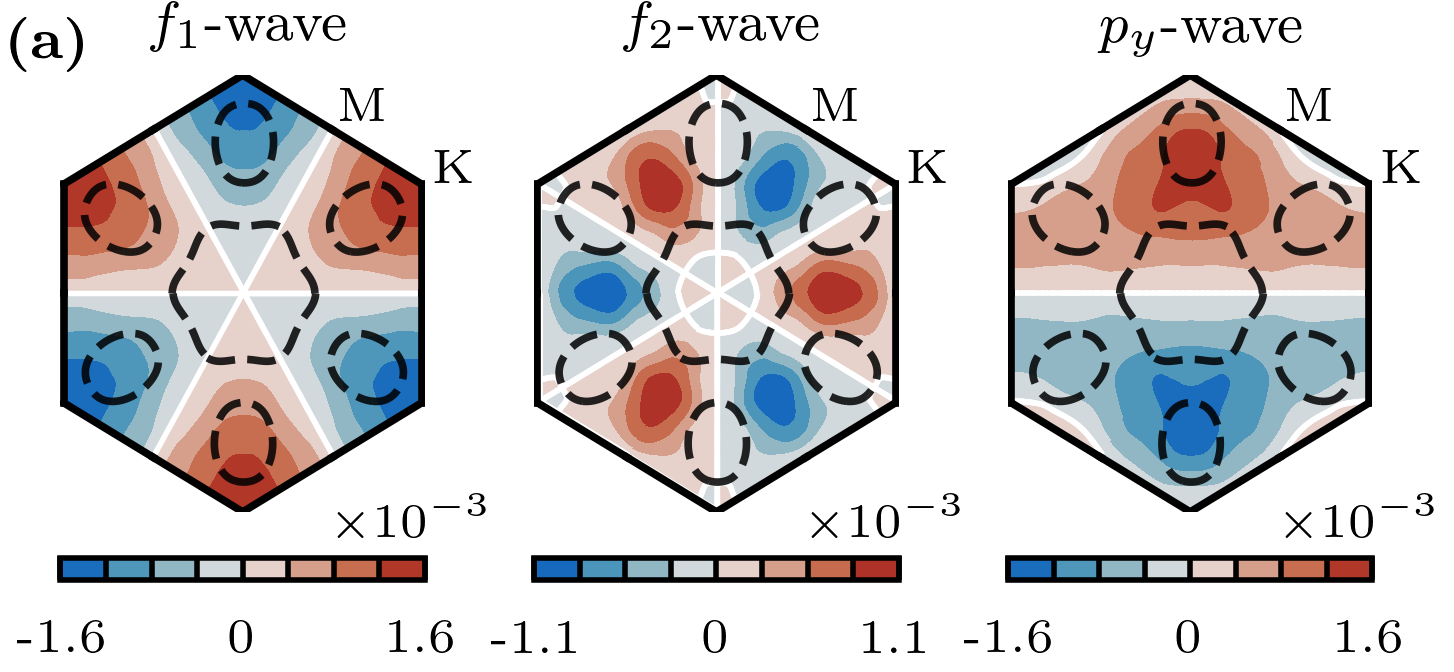}\vspace{0.8em}
	\includegraphics[width=0.48\textwidth]{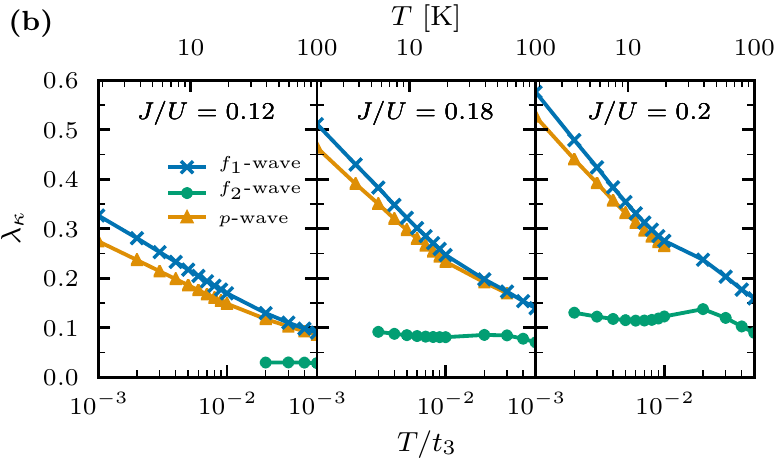}
	\caption{(a) Possible triplet-pairing symmetries of the superconducting gap. Shown is the orbital trace of the converged order parameter \(\Delta(i\omega_1,\bm{k})\) for \(T/t_3=0.003\) and \(J/U=0.2\). The line nodes (solid white) intersect differently with the Fermi surface (dashed black) depending on the gap symmetry. (b) Temperature dependence of the superconducting eigenvalue \(\lambda_{\kappa}\) for different gap symmetries  \(\kappa =f_1,\,f_2,\,p\). The panels show different exchange interactions \(J/U\). Note that the \(T\)-axis is logarithmic.}
	\label{fig:NaCoO_SC_lam_symmetry}
\end{figure}

The ferromagnetic fluctuations investigated in the previous section seem promising to mediate triplet superconductivity in \NaCoOHO. To address this question, we solve the Eliashberg equation for different pairing symmetries. Possible triplet pairings compatible with the point group of the triangular lattice are \(f_1\equiv f_{y(x^2-3y^2)}\)-, \(f_2\equiv f_{x(3x^2-y^2)}\)-, and \(p\)\,-wave, for which \(p_x\) and \(p_y\) are degenerate. The \(\bm{k}\)-dependence of the respective order parameter is depicted in Fig. \ref{fig:NaCoO_SC_lam_symmetry}(a).

The temperature dependence of the corresponding superconducting eigenvalues is shown in Fig. \ref{fig:NaCoO_SC_lam_symmetry}(b) for three different Hund's couplings. At high temperatures, the \(p\)- and \(f_1\)-wave solutions coexist with a near degeneracy that is lifted for low \(T\). There, the \(f_1\)-gap clearly shows up as the dominant pairing symmetry. Since it has line nodes between the \(\Gamma\) and M points, the \(f_1\)-gap fits well to the \(e_g\pr\) pockets of the Fermi surface  in the sense that the nodes do not intersect them. Contrarily, the \(f_2\)-gap has line nodes that intersect also the \(e_g\pr\) pockets, which explains why the \(f_2\)-symmetric gap appears unfavorable in our calculations.

While we do find an enhancement in the \(f_1\)- (dominant) and \(p\)\,-wave (subdominant) superconducting eigenvalues of the linearized Eliashberg equation upon lowering the temperature, we do not find triplet superconductivity to be realized on the order of experimental \(\Tc\). The eigenvalue of the leading \(f_1\)-symmetric gap stays below 0.6 at \(T/t_3=10^{-3}\) corresponding to approximately 2~K. Comparing \(\lambda_{f_1}\) for different \(J/U\) indicates an increase of superconducting pairing strength since the F fluctuations are enhanced. Therefore, it might be possible that the \(f_1\)-pairing is realized for larger \(J/U\)  but we cannot access this regime, since it is masked in our FLEX calculations by the magnetic instability.

\begin{figure*}[t]
	\centering
	\includegraphics[width=0.95\textwidth]{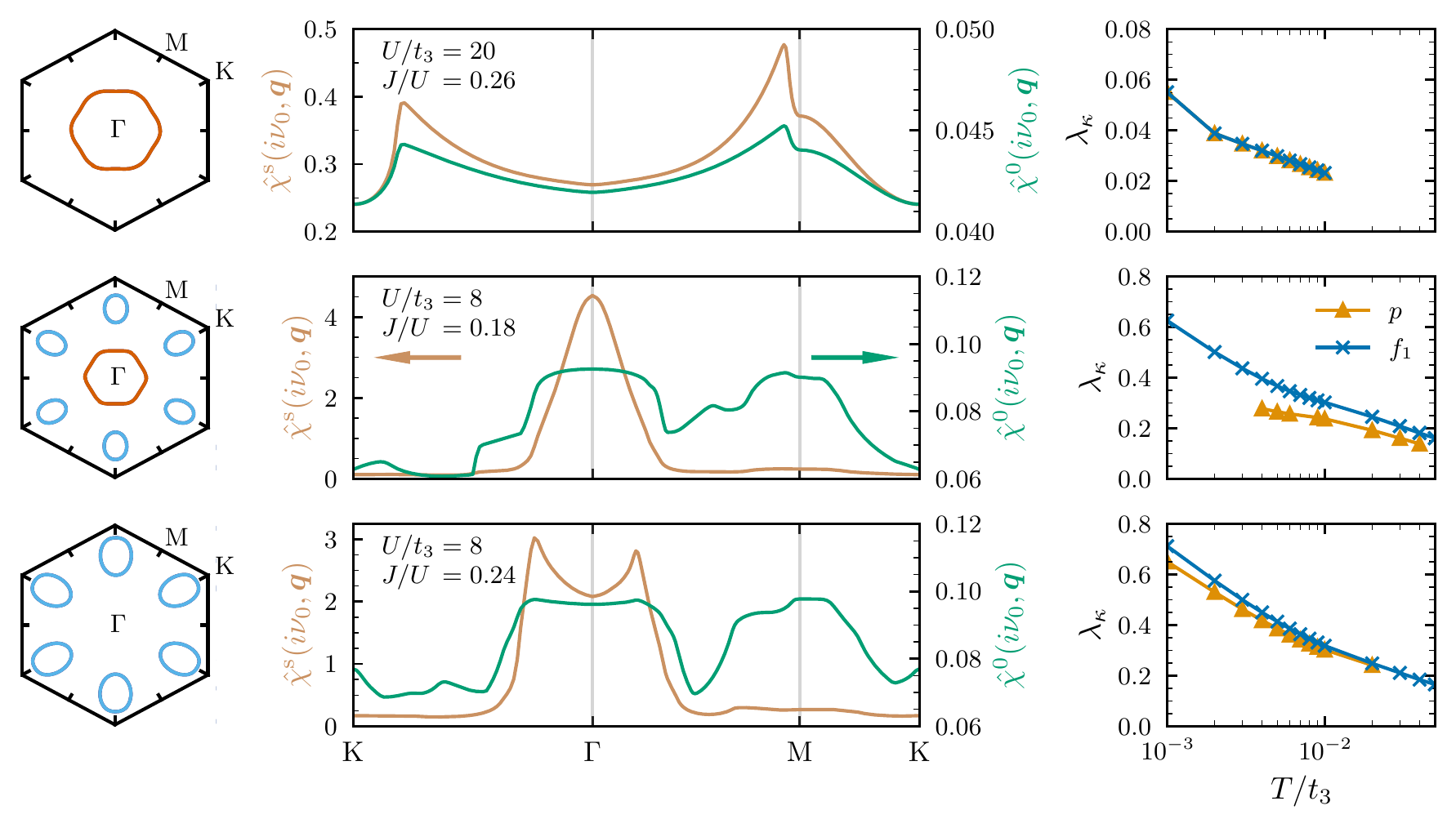}
	\caption{Dependence of magnetic fluctuations and eigenvalues of the superconducting Eliashberg equation on Fermi surface topology. Each row shows the non-interacting Fermi surface, maximal eigenvalue of irreducible and spin susceptibility, and superconducting eigenvalue at \(T/t_3=0.003\) for the maximally convergable interaction parameters. The top row corresponds to a Fermi surface composed of the \(a_{1g}\) pocket only (\(\Delta_{\mathrm{CF}}=-1.2\)), the middle row to both pocket types being present (\(\Delta_{\mathrm{CF}}=0.4\)), and the bottom row to only the \(e_g\pr\) pockets existing (\(\Delta_{\mathrm{CF}}=9.0\)).}
	\label{fig:NaCoO_FS_Study}
\end{figure*}

\subsection{Influence of the Fermi Surface}
The Fermi surface topology naturally affects the magnetic and superconducting fluctuations, whereas the exact shape of the Fermi surface for \NaCoOHO is an open question, as explained in Sec. \ref{sec:NaCoO2_structure}. Therefore, it is insightful to investigate how the magnetism and superconductivity depend on Fermi surface topology. We compare the situation with \(a_{1g}\) and \(e_g\pr\) pockets present [Fig. \ref{fig:NaCoO_system}(e)] considered so far to the cases in which either the \(a_{1g}\) or \(e_g\pr\) pockets are absent (Fig. \ref{fig:NaCoO_FS_Study}). The Fermi surface with suppressed \(e_g\pr\) pockets corresponds to the results observed in ARPES measurements \cite{Hasan2004,Yang2004,Yang2005,Shimojima2006,Geck2007}. The latter case, on the other hand, avoids any nodes of the \(f_1\)-symmetric gap on the Fermi surface which leads to the realization of \(f\)-wave superconductivity in the single-band case \cite{Kuroki2004}. 

We control the Fermi surface shape in our model via the filling \(n\) and crystal field splitting \(\Delta_{\mathrm{CF}}\). In the following, calculations we set the filling to \(n=5.6\). If we consider the Fermi surface with only the \(e_g\pr\) pockets being present, then their effective hole doping of 0.4 is equal to the hole doping of the \(e_g\pr\) pockets in our previous calculation for \(n=5.355\) and \(\Delta_{\mathrm{CF}} = 0.4\). By this, we can directly estimate the influence of neglecting the \(a_{1g}\) pocket. Furthermore, \(n=5.6\) corresponds to the \(t_{2g}\) filling reported for measurements of superconductivity when considering the additional H\(_3\)O\(^+\) doping \cite{Sakurai2015}. We choose the crystal field splitting as \(\Delta_{\mathrm{CF}}=-1.2,0.4,9.0\) to create the three different Fermi surface topologies as shown in the left column of Fig. \ref{fig:NaCoO_FS_Study}. For each, we performed calculations with different interaction parameters \mbox{\(U\) and \(J\)}.

In the remaining panels of Fig. \ref{fig:NaCoO_FS_Study}, we present \(\hat{\chi}^0\) and \(\hat{\chi}^{\mathrm{s}}\) at \(T/t_3=0.003\) and the superconducting eigenvalue \(\lambda_{\kappa}\) for the maximal values of \(U\) and \(J\) for which we were able to converge the FLEX loop. In the case of a single \(a_{1g}\) Fermi sheet, strong magnetic fluctuations do not emerge. If \(e_g\pr\) pockets exist, intra-pocket scattering strongly enhances F fluctuations. This can be seen both in the case with \(a_{1g}\) and \(e_g\pr\) pockets being present and in the case of only \(e_g\pr\) pockets existing, where we can stabilize FLEX solutions with sizable F or more generally long-wavelength spin fluctuations.

Evaluating the eigenvalues of the linearized Eliashberg equation shows that any spin-fluctuation-induced superconducting pairing is strongly suppressed in the absence of the \(e_g\pr\) pockets. Since the AF fluctuations are dominant in this scenario, we also tried to solve the Eliashberg equation for \(d\)\,-wave symmetry. However, we could not find a converged solution. If the material actually exhibits an \(a_{1g}\) Fermi surface only, a different mechanism has to be considered to explain the superconductivity. In the cases with the \(e_g\pr\) pockets present and correspondingly stronger F fluctuations, we again find the dominant \(f_1\)-wave together with subdominant \(p\)\,-wave symmetric  solutions of the linearized Eliashberg equation. By excluding the \(a_{1g}\) pocket from the Fermi surface, the superconducting pairing strength in the aforementioned \(f_1\)- and \(p\)\,-wave channels is increased, likely due to the absence of gap nodes intersecting with the Fermi surface in this case. Nonetheless, even in the absence of the \(a_{1g}\) Fermi pockets we do not find the superconducting transition on the order of experimental \(\Tc\). As previously discussed, the transition might occur for larger values of \(U\) or \(J\), which are, however, outside the region where we could stabilize the FLEX self-consistency loop.

\section{Conclusion}
We implemented the FLEX approximation using the IR basis to study magnetism and superconductivity in the Hubbard model and \NaCoOHO. Benchmark calculations on the Hubbard model showed an excellent agreement with previous FLEX calculations but at a much lower numerical cost.

This gain in numerical efficiency allowed us to turn to more realistic multi-band systems and to approach so far unexplored low-temperature regimes. We studied the dependence of magnetic and superconducting fluctuations on temperature, Fermi surface topology and interaction strength  in \NaCoOHO. We found the existence of \(e_g\pr\) pockets on the Fermi surface to be crucial, in order to generate strong ferromagnetic fluctuations. Concerning superconducting pairing, we find the \(f_{y(x^2-3y^2)}\)-wave symmetry to be dominant over other triplet-pairing symmetries at low temperatures. We do not, however, find the superconducting transition on the order of the experimental \(\Tc\), but our calculations indicate that the spin-fluctuation-driven transition takes place at significantly lower temperatures. This situation might still change for larger interactions, which are, however, inaccessible within FLEX because of too strong magnetic fluctuations. Studies employing other methods could give more insight on this question. If the \(e_g\pr\) Fermi pockets are absent, we only find weak magnetic fluctuations, which cannot establish superconductivity. In this case, the pairing mechanism has to be of a different origin. 

In summary, we have shown that the FLEX+IR approach enables the study of complex multi-orbital systems at low-temperature scales not accessible with conventional Matsubara frequency grid sampling. This should bring further systems featuring possibly an interplay of spin fluctuations and superconductivity into the reach of FLEX calculations at experimentally relevant temperature scales. Since another limiting factor of Green function methods is the momentum integration in the Brillouin-zone, a combination with e.g. adaptive \(\bm{k}\)-space sampling methods \cite{Eiguren2014,LafuenteBartolome2020,Zantout2019} could further extend the range of possible systems. Interesting grounds to be explored range from moir\'{e} superlattice systems to realistic multi-band models of infinite-layer nickelate compounds.

\begin{acknowledgments}
NW thanks the University of Tokyo for hospitality during his research stay, where ideas presented in this work were conceived. N.W. and T.W. acknowledge support by the Deutsche Forschungsgemeinschaft (DFG) via RTG 2247 (QM3) (Project number 286518848) and European Commission via the Graphene Flagship Core Project 3 (Grant agreement ID: 881603). E.v.L. is supported by the Central Research Development Fund of the University of Bremen. The authors acknowledge the North-German Supercomputing Alliance (HLRN) for providing computing resources that have contributed to the research results reported in this paper.
\end{acknowledgments}

\appendix
\section{Tight-binding model for \NaCoOHO}\label{sec:app_NaxCoO_TB}
The tight-binding model to describe the electronic structure of \NaCoOHO is constructed following Ref. [\onlinecite{Mochizuki2005}] and reads:
\begin{align}
	H_{\mathrm{TB}} =  \sum_{\gamma,\gamma\pr,\sigma}\qty(\varepsilon_{\bm{k}}^{\gamma\gamma\pr}+\frac{\Delta_{\mathrm{CF}}}{3}(1-\delta_{\gamma\gamma\pr}))\cdag{\bm{k}\gamma\sigma}\cno{\bm{k}\gamma\pr\sigma}
\end{align}
Here, the summation goes over spin \(\sigma\) and the \(d\)-orbitals \(\gamma\) of the \(t_{2g}\) manifold. The first term describes the kinetic energy and the second term includes the crystal electric field \(\Delta_{\mathrm{CF}}\) due to the trigonal distortion of the CoO\(_6\) octahedra (cf. Fig. \ref{fig:NaCoO_system}(c)). The band dispersion is given by
\begin{align*}
\begin{split}
	\varepsilon_{\bm{k}}^{\gamma\gamma} &= 2t_1\cos \bm{k}_{\alpha}^{\gamma\gamma} + 2t_2\qty[\cos \bm{k}_{\beta}^{\gamma\gamma} + \cos\qty(\bm{k}_{\alpha}^{\gamma\gamma} + \bm{k}_{\beta}^{\gamma\gamma})]\\
	&\qquad+ 2t_4\qty[\cos(2\bm{k}_{\alpha}^{\gamma\gamma} + \bm{k}_{\beta}^{\gamma\gamma}) + \cos(\bm{k}_{\alpha}^{\gamma\gamma} - \bm{k}_{\beta}^{\gamma\gamma})]\\
	&+ 2t_5\cos(2\bm{k}_{\alpha}^{\gamma\gamma})\quad,
\end{split}\\
\begin{split}
	\varepsilon_{\bm{k}}^{\gamma\gamma\pr} &= 2t_3\cos \bm{k}_{\beta}^{\gamma\gamma\pr} + 2t_6\cos \bm{k}_{\beta}^{\gamma\gamma\pr} + 2t_7\cos(\bm{k}_{\alpha}^{\gamma\gamma\pr} + 2\bm{k}_{\beta}^{\gamma\gamma\pr})\\
	&\qquad + 2t_8\cos(\bm{k}_{\alpha}^{\gamma\gamma} - \bm{k}_{\beta}^{\gamma\gamma}) + 2t_9\cos(2\bm{k}_{\alpha}^{\gamma\gamma} + \bm{k}_{\beta}^{\gamma\gamma})
\end{split}
\end{align*}
where \(\bm{k}_{\alpha}^{xy,xy} = \bm{k}_{\alpha}^{xy,zx} = \bm{k}_1\), \(\bm{k}_{\beta}^{xy,xy} = \bm{k}_{\beta}^{xy,zx} = \bm{k}_2\) , \(\bm{k}_{\alpha}^{yz,yz} = \bm{k}_{\alpha}^{xy,yz} = \bm{k}_2\), \(\bm{k}_{\beta}^{yz,yz} = \bm{k}_{\beta}^{xy,yz} = -(\bm{k}_1+\bm{k}_2)\), \(\bm{k}_{\alpha}^{zx,zx} = \bm{k}_{\alpha}^{yz,zx} = -(\bm{k}_1+\bm{k}_2)\) and \(\bm{k}_{\beta}^{zx,zx} = \bm{k}_{\beta}^{yz,zx} = \bm{k}_1\) with \(\bm{k}_1\) and \(\bm{k}_2\) being the reciprocal lattice vectors defined by the triangular lattice in Fig. \ref{fig:NaCoO_system}(a).

We employ the hopping parameters \(t_1 = 0.45,\, t_2 = 0.05,\, t_3 = 1,\, t_4 = 0.2,\, t_5 = -0.15,\, t_6 = -0.05,\, t_7 = 0.12,\, t_8=0.12\) and \(t_9 = -0.45 \) where \(t_3\) is the unit of energy. Setting \(\Delta_{\mathrm{CF}}=0.4\) and \(t_3 \equiv 0.18\)~eV reproduces LDA band structure calculations \cite{Singh2000} well, particularly around the Fermi level. The value of \(\Delta_{\mathrm{CF}}\) significantly influences the Fermi surface topology.

\bibliography{references}

\end{document}